# Bloch oscillations and subwavelength focusing in stacked fishnet metamaterials

Xiaopeng Su,[1, 2] Zhijie Gong,[1] Hongrui Wu,[3] Yanting Lin,[2] Zeyong Wei,[1$] Chao Wu,[1] & Hongqiang Li[1, 2+]

[1] School of Physical Science and Engineering, Tongji University, 200092, Shanghai, China

[2] The Institute of Dongguan-Tongji University, 523808, Dongguan, Guangdong, China

[3] Department of Physics, University of California San Diego, 92093, CA, USA

To whom correspondence should be addressed. E-mail: weizeyong@163.com; li@idtu.cn

In this letter, we introduce stacked fishnet metamaterial for steering light in microwave region. We numerically demonstrate that optical Bloch oscillations and a focus of as small as one sixth of a wavelength can be achieved. The flexibility of varying geometrical parameters of the fishnet slabs provides an efficient way for tuning its local effective media parameters and opens the possibility for controlling light arbitrarily. The experiment verifies subwavelength-sized light focusing effect by scanning magnetic field at the surface of the sample directly.

The dynamics of waves in artificial structures is a common and fundamental issue in many branches of physics, such as condensed matter physics, acoustics, quantum mechanics or photonics. Different physical systems are possible to map to each other due to the similarity in the mathematical form of corresponding equations. One of classical analogy among them is reproduction of behavior of electrons in a condensed matter or quantum system by photons in an equivalent optical system. As the rapid development of modern nano-fabrication as well as advanced photonic detectors, propagations of lights in nano-structures are accessible to physicists to observe and detect in a more visual way. Bloch oscillations, Zener tunneling , spin Hall effect [1-3] or Bose-Einstein condensation[4], as dynamic behaviors of electrons for condensed matter phenomena, can be visualized in artificial photonic systems such as waveguide arrays, photonic crystals, metamaterials or optical microcavity. Artificially designing a mimic environment of solid state by periodic or aperiodic structures at subwavelength scale is considered as an efficient way to routing the flow of light. Under such kind of concept, metamaterial, comprised of subwavelength artificially-sized metallic resonant units, is well studied to realize the precise control over the propagation of electromagnetic wave. As a result, extraordinary optical effects as well as mimicking of condensed matter phenomena, such as negative refraction, perfect lens, cloaking Bloch Oscillations and spin Hall effect [2, 3] are successively reported in photonic systems.

One of typical example of a vivid mapping of electronic wave packet to photons in classical optical system is Bloch oscillations, which is firstly addressed by Felix Bloch about 80 years ago. Under the influence of a constant external field, a quantum particle in tilted periodic potential is accelerated until its Bloch momentum reached the edge of Brillouin zone, then it undergoes Bragg reflection and deceleration in the external field. The particle subsequently returns to the initial condition and starts the next period of motion as before. The prediction of this fantastic phenomenon is suffered controversies for a long time as it is rarely observed directly in regular crystals. The coherence time of electronic wave packet is much shorter in comparison with the period of Bloch oscillations, [5, 6]. However, in semiconductor superlattices systems[7], the much faster oscillation period give the opportunity to observe Bloch oscillations. On the other hand, in atomic systems[5, 8] Bloch oscillations can be also observed in optical lattices for quite long lifetime of the atoms. Another important classical optical system for mapping is waveguide arrays, which pave the way to achieve visualization of propagations of lights. In waveguide arrays, keeping the original physical mechanism, the behaviors of electronic wave packets in time domain is vividly presented in the form of spatial light intensity distributions which are facilitate to be observed. Dynamical behaviors of lights, such as discrete diffraction Bloch oscillations, Zener tunneling, discrete solitons, are achieved in waveguide arrays. Evanescent coupling and tunneling between adjacent waveguides bridge lights from one channel to another one and manage its dispersion and way of propagation.

In this paper, we demonstrate that optical Bloch oscillations and subwavelength focusing in a system of waveguide arrays that consist of alternative dielectric layers and metallic fishnet structures. The environment of a crystal influenced by a static electric field is mimicked in a system of plasmonic metamaterial. In comparison with metal dielectric waveguide arrays, primarily considering the coupling between adjacent waveguides, stacked fishnet waveguide arrays (SFWAs) demonstrated in this paper provide an additional freedom in tailoring of plasmonics due to the mimicking of surface plasmon polaritons (SPPs) by structured surface.

Instead of gradient of dielectric layers, i. e, permittivity, thickness and temperature, across the entire arrays, the process of mimicking equivalent potentials in our classical optical system is dictated by metallic fishnet perforated with different size of holes. As a consequence, the properties of dispersion of periodic holes in a layer as well as mutual coupling between adjacent channels sandwiched by metallic fishnets are deliberately altered with the variety of geometric parameters of holes. Moreover, subwavelength focusing effect is also realized by tailoring plasmonics under precise design. Most power of incidence could be concentrated into a channel with deep subwavelength scale. Propagation of electromagnetic wave through SFWAs shows negative refraction for oblique incidence. We prepare a three-dimensional metamaterial of SFWAs as sample to demonstrate the experimental observation of subwavelength focusing by scanning field intensity distributions. Optical Bloch oscillations and subwavelength focusing will be sequentially illustrated in the following sections of this paper.

## Results

**Spoof plasmonic waveguide system.** We observe the optical Bloch oscillations in the stacked fishnet structures shown in Fig. 1a. Spatial Bloch oscillations of SPPs or spoof SPPs in metal waveguide arrays (MWGAs) have been observed in the previous studies[6, 9-12]. However, to the best of our knowledge, optical Bloch oscillations in waveguide arrays reported until now are mostly realized by refractive index gradient of dielectric slabs across the MWGAs. Here, we realize optical Bloch oscillations by tuning the geometrical parameters of holes in local fishnet arrays, which can offer an effective refractive index gradient across the arrays in an indirect way. However the index of dielectric slabs and the period of waveguide arrays are keeping constant. The local geometrical parameters of unit cells in a certain metallic fishnet layer are unchanged.

Schematic pictures of the system under study are shown in the Fig. 1a. The structure is composed of 30 arrays of dielectric slabs and 29 arrays of metallic layers with different size of holes drilled in them. The period of structure in *YZ* plane is *d*=6mm. The thickness of dielectric slabs is $h_1$=0.5mm and the thickness of metallic layers is $h_2$=1mm, so the period of waveguides is

$p$=1.5mm. All the parameters stated here guarantee the unit cell sizes in our structure are far below the operational wavelength that we care about. The dielectric constant of the dielectric slabs is $\varepsilon_d$=2.65. The length in $Z$ direction is 360mm (60 lattices). A TM-polarized (polarized in the $Z$ direction) Gaussian beam is launched into the stacked fishnet waveguide arrays from the boundary with the magnetic field amplitude profile $H_z = \exp[-(x-x_0)^2/w_0^2]$, where $x_0$=24.25mm (the center position in $X$ direction of 17th dielectric slab) and $w_0$=8mm. They are the center position and the full width at half maximum (FWHM) of Gaussian beam, respectively. The incident beam frequency is 11.5GHz, which is far below the cut-off frequency of holes drilled in structures. The length of rectangle holes in $Y$ direction and $Z$ direction are $b_i \times a_i$, and their local parameters have been shown in Fig. 1b with red scatters. To observe spatial Bloch oscillations in our system, here, we set elaborate design for local drilled holes sizes in order to obtain the gradient in effective relative permittivity. The real part of effective relative permittivity parameters for each fishnet slabs (blue star-line) are close to those we expected (blue rhombus-line). These effective electromagnetic parameters of each metallic fishnet layers are obtained by S-parameter retrieval method[13]. Effective electromagnetic parameters are sensitive to local unit geometrical size. The calculated results in Fig. 1b show that PEC layers drilled with variable size of holes perform different electromagnetic properties in different position. The operational frequency is far below the cutoff frequency of fishnet metamaterial. The allowed waveguide modes in the biggest drilled holes for fishnet slab at the position $x$=43mm own the lowest cutoff frequency at $f_{c\min} = \frac{c_0}{2a_{\max}} \approx 25.8GHz$, where $c_0$ is the light velocity in vacuum. Other drilled holes with relatively smaller sizes support waveguides modes at higher frequency. So the designed sizes are subwavelength scale ones considering our operational wavelength.

In the long wavelength limit, metallic surface perforated with holes can be homogenized into a single-negative medium with electric response in the form of Drude type model[14]. So it is reasonable for us to consider this stacked fishnet arrays as a kind of spoof plasmonic waveguide arrays (SPWAs). Since the variety of size in each fishnet layers is not severe, the SPWAs can be thought of approximately as locally uniform. For the TM-polarized wave, we can get the Bloch

wave dispersion relation by transfer matrix method at a certain frequency[15]:

$$\cos[k_x(h_1+h_2)] = \cos(k_1 h_1)\cos(k_2 h_2) - (\frac{\varepsilon_m^2 k_1^2 + \varepsilon_d^2 k_2^2}{2\varepsilon_m \varepsilon_d k_1 k_2})\sin(k_1 h_1)\sin(k_2 h_2) \qquad (1)$$

in which $k_x$ is the Bloch wave vector, $k_1 = \sqrt{\varepsilon_d k_0^2 - \beta^2}$ and $k_2 = \sqrt{\varepsilon_m k_0^2 - \beta^2}$; $\beta$ is the propagation wave vector along the Y direction; $k_0$ is the propagation wave vector of incident light in the free space. $h_1$ and $h_2$ are the thickness of a dielectric slabs and metallic layers, respectively. $\varepsilon_m$ is the local effective relative permittivity of the perforated metallic layers, and $\varepsilon_d$ is the relative permittivity of the dielectric slabs.

**Bloch oscillations in stacked fishnet metamaterials.** Fig. 2a plots the simulated magnetic field distributions $H_z$ in the XY plane. In our finite-difference time-domain (FDTD) simulations, a Gaussian beam is normally incident to the model. Periodic boundaries are set in the Z direction representing the repeating units. The ray initially moves to the left part of structure and reaches the leftmost position. Then it moves to the right and returns to the origin position in X direction. Finally, the magnetic field distribution is almost recovered after a period propagation. The trajectory of light is similar to the behavior of optical Bloch oscillations result from refraction index, width or even temperature gradient in dielectric arrays[6, 9-11, 16]. The period of ray trajectory in the stacked fishnet structure is 310mm. Fig. 2b plots the calculated dispersion in the X - $k_x p$ space. The region in blue corresponds to the different values between two sides of equation (1) approaching to zero, which predict the movement direction of ray trajectory. The beam originates at the position $X=x_0$ corresponding to the center of the first Brillouin zone ($x_0$, 0) and moves in the negative X direction with the increasing $k_x$ until it reaches the boundary of first Brillouin zone ($k_x p=\pi$). Then the beam undergoes Bragg reflection and appears at the other edge of the first Brillouin zone ($k_x p=-\pi$). Once the contour has crossed the boundary of Brillouin zone, it moves to the positive X direction as $k_x$ towards zero (the center of the first Brillouin zone). Finally the contour returns to origin position ($x_0$, 0) completing a whole Bloch oscillation period. The optical Bloch oscillations come from the alternating total internal reflection and Bragg reflection between the boundaries of waveguide arrays[6, 9, 10]. The variety of local geometrical parameters plays the similar role with the refraction index gradient or waveguide width gradient across the structure and devotes to spatial Bloch oscillations. All of them result in propagation constant variety of wave across arrays, which can be solved by equation (1). This optical Bloch oscillation is the spoof SPPs one, and it is an analogue of the one in electronic system in a crystal[17]. Our exact design in fishnet unit cells mimic the function of external static electric field influencing on a crystal[10].

**Subwavelength focusing effect in stacked fishnet metamaterals.** For the inhomogeneous system demonstrated here, we have observed optical Bloch oscillations in a unidirectional geometric gradient variety case. In the following section, an inhomogeneous system with symmetry in geometric gradient variety is under study. A similar stacked fishnet model is investigated to discover the possibility for realization subwavelength focusing. Schematic picture is shown in the inset a of Fig. 3, and inset b displays the photo of probe for Magnetic field measurement. Our structure shown in Fig. 3 for wave focusing is composed of 28 perforated PEC layers that alternate with 27 dielectric slabs. The length of rectangle hole in $Y$ direction is $b=2$mm, and the length of each rectangle hole in $Z$ direction is $a_i$ (mm) which varies with the position of $X$ corresponding to different PEC layers. The system is a symmetrical one of $YZ$ surface (as shown in inset a). Thus the cut-off frequencies of air holes vary with the position. The period of structure in $YZ$ plane is still $d=6$mm. For facility of fabrication, the thickness of dielectric slabs is $h_1=1.5$mm and the thickness of metallic layers is $h_2=0.1$mm, so the period of waveguide is $p=1.6$mm. The dielectric constant of the dielectric slabs is $\varepsilon_d=2.65$. The total length in the $Y$ direction is 210mm (35 lattices). The TM-polarized (polarized in the $Z$ direction) Gaussian beam with 75mm waist radius is launched into the stacked fishnet arrays.

Fig. 4a plots the simulated results of the magnetic field intensity for stacked fishnet structure in Fig. 3 at the frequency of 10.5GHz. Fig. 4b plots results of the magnetic field intensity for the homogenized media waveguide arrays at the same frequency. These effective local electromagnetic parameters are still obtained by S-parameter retrieval method. The two magnetic field intensity distributions are similar to each other with the significant focusing spot around

$Y$=200mm. In our FDTD simulations, a Gaussian wave in $XY$ plane is normally incident to the model. Periodic boundaries are set in the $Z$ direction representing the repeating units. The focusing effect is clearly shown in Fig. 4a, and the focal length is about 198mm occurring over a length scale that corresponds to several wavelengths. Fig. 4c depicts the intensity distribution profile in the focal plane ($Y$=198mm). The intensity peak in Fig. 4c is about 20 times larger than the incident pulse intensity. Full width at half-maximum (FWHM) intensity is 4.7mm (the total width of three dielectric slabs and two PEC layers). Thus the fields are mainly confined in the central dielectric gaps between the metallic fishnet films. For electric field intensity distribution, its maximum intensity is located at the surface region of metallic films implying that spoof SPPs modes excited by the formation of resonance[14, 18-21].

We note that the effective parameters of homogenized media are tuned by the local geometrical parameters of perforated metallic slabs which vary in the $X$ direction. So the local propagation wave vectors in a certain uniform waveguide system can be solved by equation (1). Importantly they are all the functions of position $X$ because of geometrical parameters variety along $X$. So, the dispersion relation of the system takes the form as[22]:

$$k_y(x) = \frac{\beta_s(x)+\beta_a(x)}{2} + \frac{\beta_s(x)-\beta_a(x)}{2}\cos(k_x p) \qquad (2)$$

Where $k_y$ and $k_x$ are the vector components along $Y$ and $X$ directions, $\beta_s$ and $\beta_a$ are the symmetry and anti-symmetry mode's propagation wave vector of $\beta$, and they are solved by equation (1) for $k_x$=0 and $k_x=\pi/(h_1+h_2)$, respectively. In the uniform case, previous works have shown that negative refraction can be achieved because of the anomalous coupling[22, 23].

Even in our inhomogeneous stacked fishnet case, the local $\beta_a$ is bigger than local $\beta_s$, which means the coupling constant between the neighboring arrays is negative. Fig. 4d shows a wave at the frequency of 10.5GHz is incident on the structure with the incident angle of 30 degrees. The focus spot ($X$<0) moves away from the direction in which the wave is tilted ($k_x$>0). The yellow arrows show the direction of incident wave vector. It clearly shows that the oblique incident wave in the structure suffer the progress of the negative refraction, which comes from the negative evanescent coupling between the adjacent waveguides[10, 24-26].

In the Fig. 5a and b, we plot the local dispersion relation of our structure with several certain constant positions *X*. All dispersion curves in the central part of first Brillouin zone are hyperbolic like curves[27]. The arrows above the dashed curves in Fig. 5 demonstrate the direction of group velocity, and they are always normal to the equi-frequency contour. $k_y$ is a conserved quantity because of the uniformity along the direction *Y*. We can see from Fig. 5a that for certain $k_y$ the local dispersion relation of our structure in different positions *X* suffered with different lateral components. These lateral components come from the inhomogeneous arrangement in the structure and exert influence on the direction of lights during propagation. So lights in the structure may change their propagating direction despite of normal incidence. The gradual changes in the geometrical parameters across the system create the dispersion curves with different curvature. The curvatures of dispersion curves for relatively closing to center arrays (i.e. blue dashed curve) are smaller than those near the margin arrays (i.e. red dashed curve). This means lights near the margin arrays are modulated by the structure more severely than those near the center arrays. Thus, the differences of these dispersion curves at different positions demonstrate that lights in the structure may change their states during progress of propagation. For oblique incident case, the group velocity of incident lights with original lateral components may suffer negative refraction because of anomalous coupling as shown in Fig. 4d. Fig. 5b shows that the directions of the refracted group velocity undergo negative refraction in the structure.

**Experiments on subwavelength focusing effect.** For the facility in fabrication of our sample by printed-circuit-board (PCB) technology, we prepared a three dimension sample to experimentally verify the subwavelength focusing effect in the stacked graded fishnet metamaterial. The experiment sample is composed by arrays of PCB slabs as shown in Fig. 3. The probe for magnetic field scanning is a ring antenna made by coaxial cable. Its diameter is about 5mm as shown in the inset b of Fig. 3. The antenna on a moveable platform is connected to a Network Analyzer and measures the complex transmission coefficient $\tilde{S}_{21}$. The measured magnetic field intensity $|\tilde{S}_{21}|^2$ at 1mm above the surface of the sample in *XY* plane is plotted in Fig. 6a. We can see that the magnetic fields intensity focus effect at the position *Y*=197mm at the frequency of 10.6GHz. The FWHM of experiment result is 5.27 mm (about $\lambda/5.4$), as shown in Fig. 6b, which has achieved the subwavelength focusing condition. We also compare the magnetic field intensity at the same position with or without our sample in Fig. 6b (the gray dot and dash lines). The magnetic field intensity with sample is about 10 times larger than that without sample at 10.6GHz. It also implies that the focusing effect at 10.6GHz actually exists because of the high signal noise ratio.

**Discussion**

The FWHM of numerical simulations is 4.7mm, which own a relatively higher resolution than experiment result. The frequency is 100MHz higher than that in simulation. These experimental errors come from materials such as the dielectric constant and the thickness of substrate. Considering the dimension of coaxial cable is about 2mm, we have to prepare an antenna whose diameter is not smaller than 5mm to guarantee the linear TM polarization wave for scanning. So the diameter of the ring antenna used to scan the magnetic field is about 5mm which is also the limit of resolution for our experiment. Hence the measured FWHM of a focusing spot is hard to be smaller than 5mm. On the other hand, we measure the magnetic field intensity in the surface 1mm over our sample, so the fields inside the sample may undergo dispersion once entering the free space. Considering the significant difference of magnetic field intensity between sample existing and free space at the same place, there is exactly field focusing effect caused by sample. For the reasons given above, we conclude that spatial resolution of focusing spot in our experiment has overcome the diffraction limit.

In conclusion, we observe optical Bloch Oscillations and experimentally verify deep subwavelength focusing effect by stacked fishnet metamaterial with varying local geometrical parameters. In long wavelength limit, metallic surface with holes on surface can support the spoof SPPs. The variety of local geometrical parameters of holes provide effective gradient of refraction index in the direction *X*. Both of them contribute to optical Bloch oscillations in microwave region. Moreover, subwavelength focusing effect is numerical calculated and experimentally observed.

We verify that the focal spot can be limited to about λ/5.4 even at the surface 1mm away from the sample, implying that a smaller spot shall exist inside it.

**Methods**

**Sample fabrication.** We prepared a three dimension sample by PCB technology to experimentally verify the subwavelength focusing effect in the stacked graded fishnet metamaterial. The pattern of copper layer that could be accurately controlled is printed on a F4BM dielectric substrate. The relative dielectric constant of the dielectric in PCB slab is 2.65±2% at the frequency of 10GHz. The thickness of each dielectric slab is in the range from 1.55mm to 1.57mm, and the thickness of a copper layer is 3oz (its thickness is 0.105mm). All the materials used in sample fabrication are as close as possible to those in simulations. Our experiment sample for wave focusing is composed of 28 perforated copper layers that alternate with 27 dielectric slabs. All PCB slabs with different size of holes drilled in copper layers are stacked in arrays according to our design order. The scale of a PCB slab in the *Y* direction is 210mm (35 units), and its scale in the *Z* direction is 102mm (17 units).

**Magnetic field measurement.** Magnetic field is measured by a small ring antenna. The polarization direction of magnetic field is perpendicular to the surface of ring. We prepare the ring antenna with a coaxial cable, and its diameter is about 5mm as shown in the inset b of Fig. 3. The antenna is connected with an Aglient N5222A Network Analyzer to measure the complex transmission coefficient $\tilde{S}_{21}$. It is installed on a movable platform which is controlled by a computer. The minimal steps of the moveable platform in the three directions (*X, Y, Z* direction) are all 1mm. A horn antenna creates a wide waist Gaussian beam which can be approximately considered as plane wave at the position several wavelengths away from the sample. During the experimental measurements, the ring antenna scans the magnetic field at the surface of sample. To avoid touching the sample and considering the size of coaxial cable, our scanning plane is set to 1mm above the surface of sample. Signals at certain position are measured and recorded by Network Analyzer after 16times average. Then, the probe moves to next position and starts the measurements again. So the magnetic field distribution at the surface of sample is mapped.

**Fig. 1**

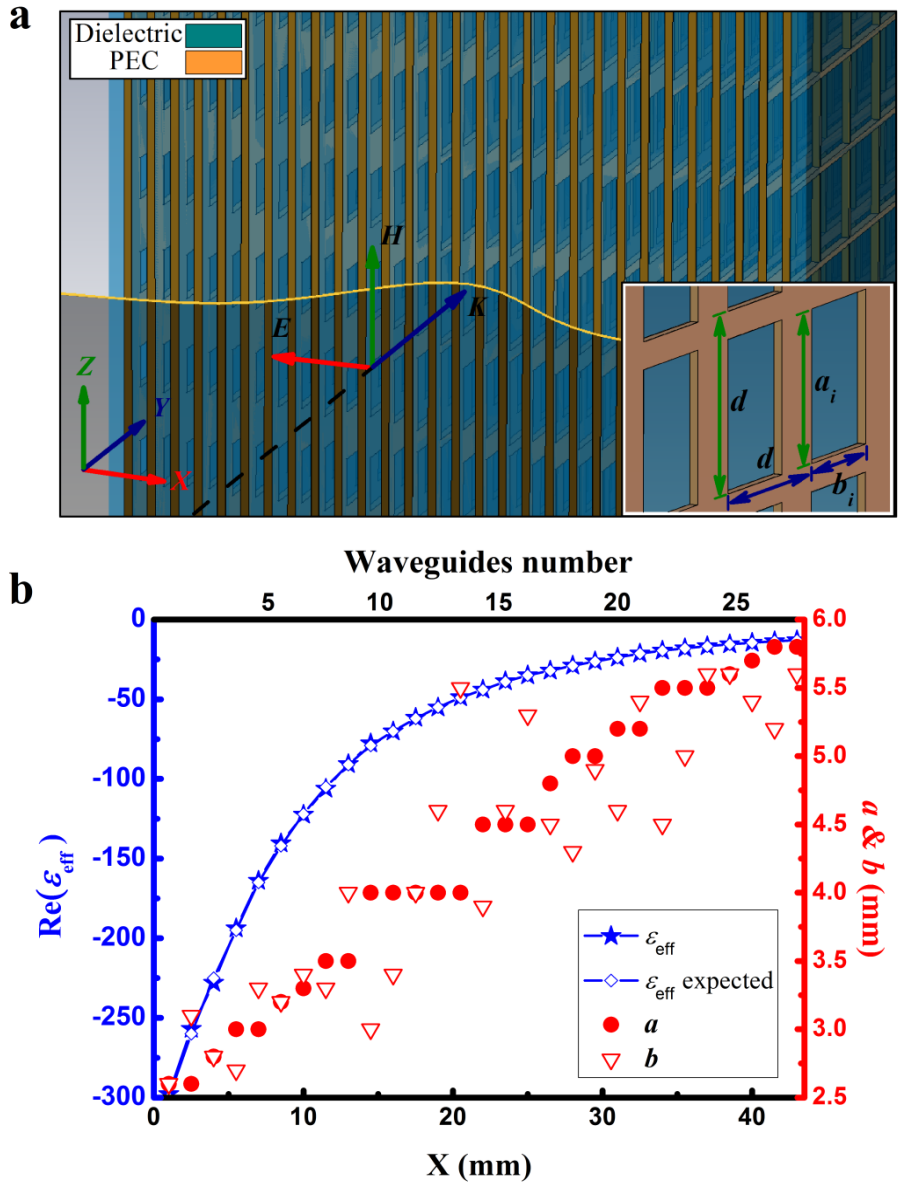

**Figure 1 | Stacked fishnet metamaterial for electromagnetic Bloch oscillations.** (a) Scheme view of dielectric slabs (cyan) and graded PEC fishnet layers drilled with rectangular holes (deep yellow) in different sizes which are gradient along the x direction. A TM-polarized Gaussian beam is launched into the stacked fishnet waveguide arrays along the y direction. The inset shows the local geometrical parameters $a_i$ and $b_i$, period $d$ of the i[th] fishnet layer. (b) the real part of effective relative permittivity Re(), extracted by S-parameter method (rhombus) and by the gradient of expected wavevector (asteroids), $a_i$ and $b_i$ (triangles and dots) and as a function of position X at i[th] fishnet layer.

Fig. 2

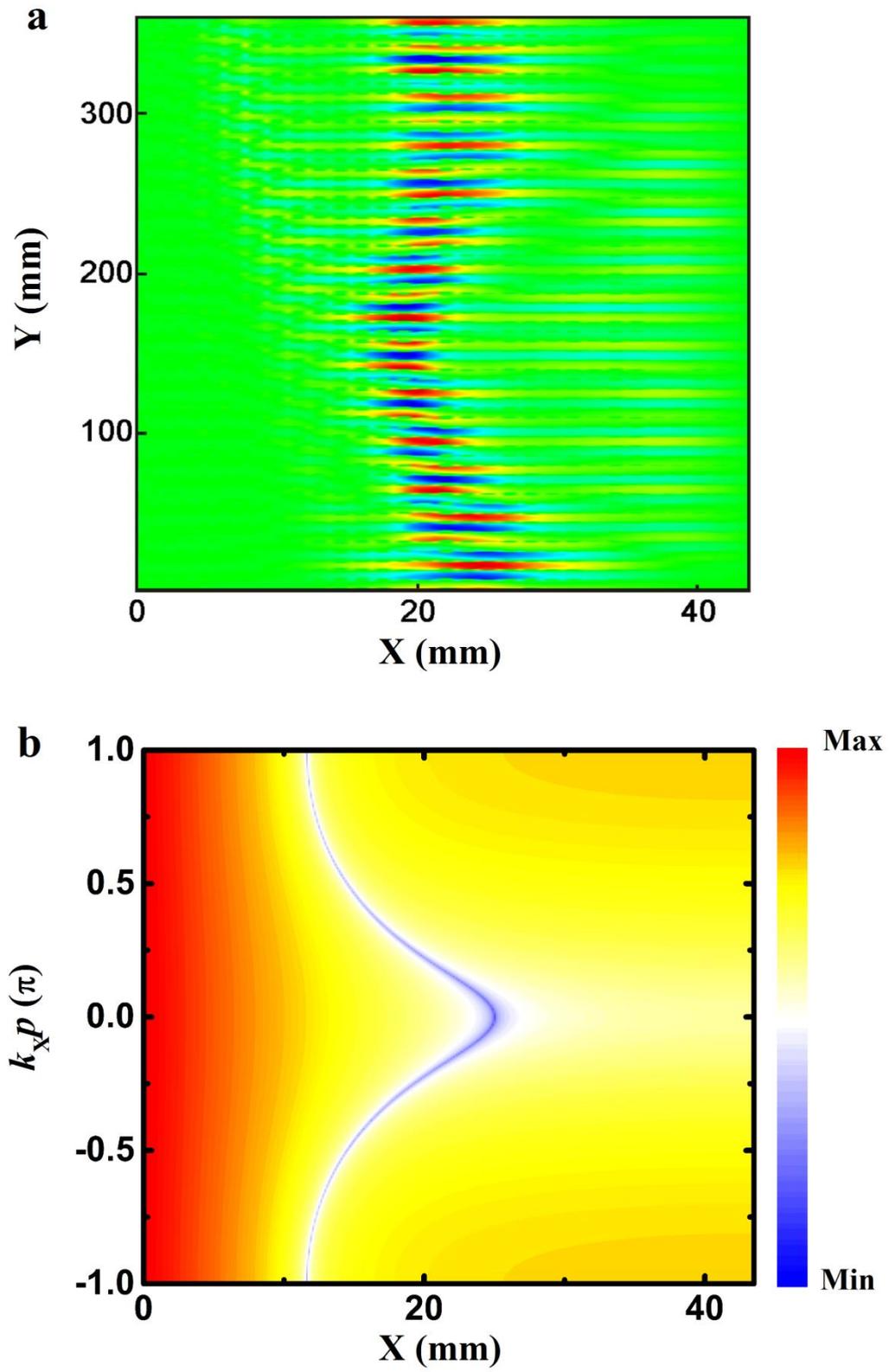

**Figure 2 | Electromagnetic Bloch oscillations in stacked fishnet metamaterial.** (a) The real part of magnetic field distribution Re($H_z$) in x-y plane within the structure as shown in Fig. 1, calculated by FDTD numerical simulations. (b) Calculated effective wavevector varied as a function of X coordinates with $\beta = 687 m^{-1}$.

**Fig. 3**

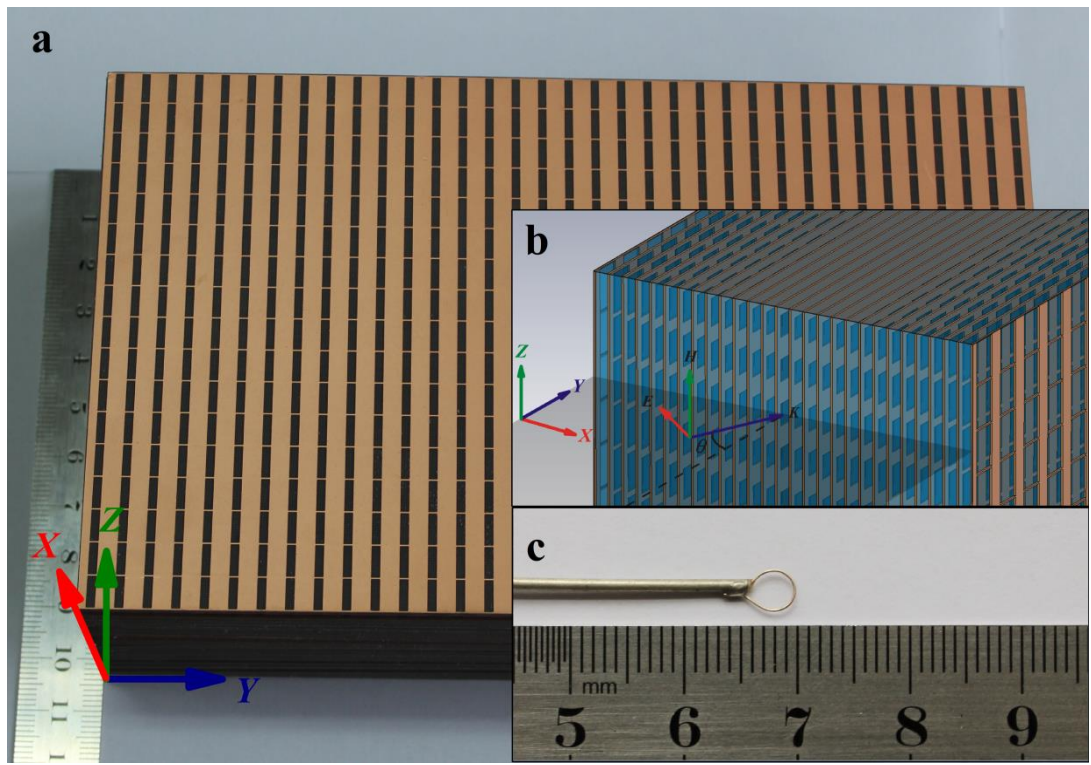

**Figure 3 | Stacked fishnet metamaterial for subwavelength focusing.** (a) Photo of our fishnet sample; (b) schematic of the sample with varied local parameters $a_i$ being symmetric about X=0 ; (c) photo of a ring probe for near-field detection. The diameter of the antenna ring is about 5mm. inset (a) is the scheme view of sample, dielectric slabs (cyan) and graded PEC fishnet layers (deep yellow) with varying width $a_i$ (from the center to the side $a_{1-14}$ =4.0, 4.1, 4.2, 4.3, 4.4, 4.5, 4.6, 4.7, 4.8, 4.9, 5.1, 5.3, 5.5, 5.8), and it is a symmetrical structure of $YZ$ surface. Inset (b) the antenna probe for magnetic field intensity measurement.

**Fig. 4**

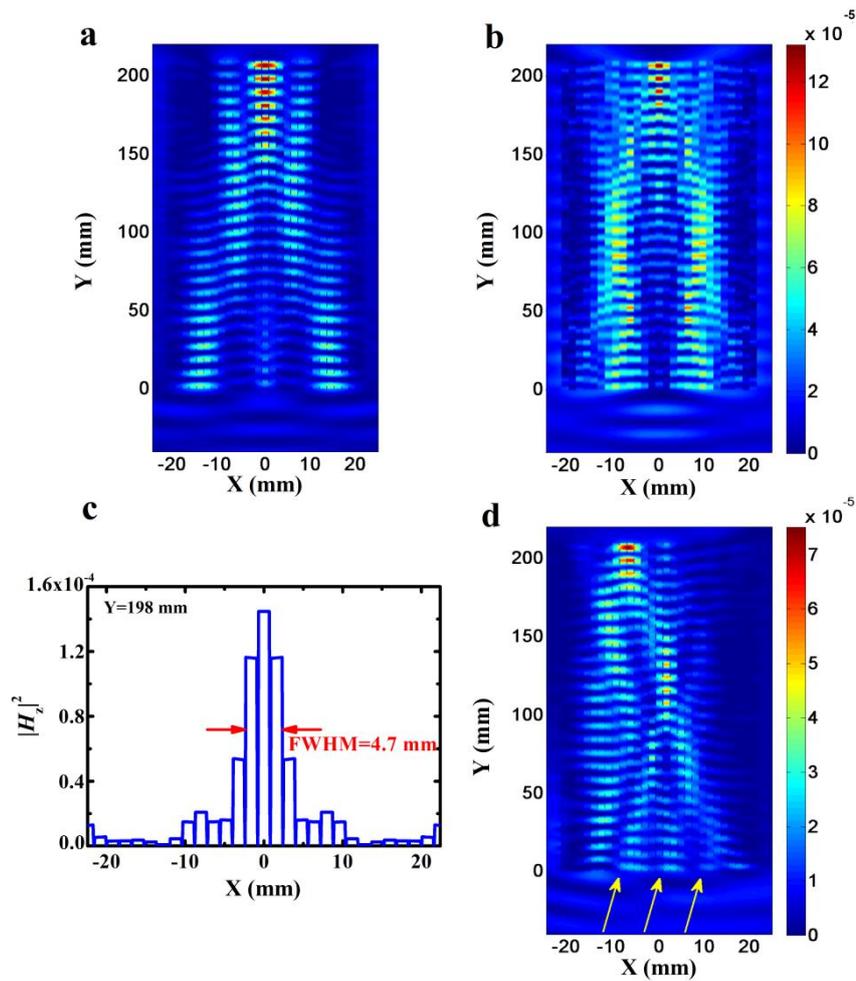

**Figure 4 | Magnetic Field Intensity $|H_z(x,y)|^2$ of subwavelength focusing.** (a) FDTD simulations at 10.5GHz under normal incidence in the structure shown in Fig. 3. (b) Calculated results for a homogenized model of the structure. (c) FWHM of focal spot at $Y = 198 mm$ shown in (a). (d) FDTD simulations at 10.5GHz at an incident angle of 30 degrees. The yellow arrows show the direction of incident wave vector outside the structure.

**Fig. 5**

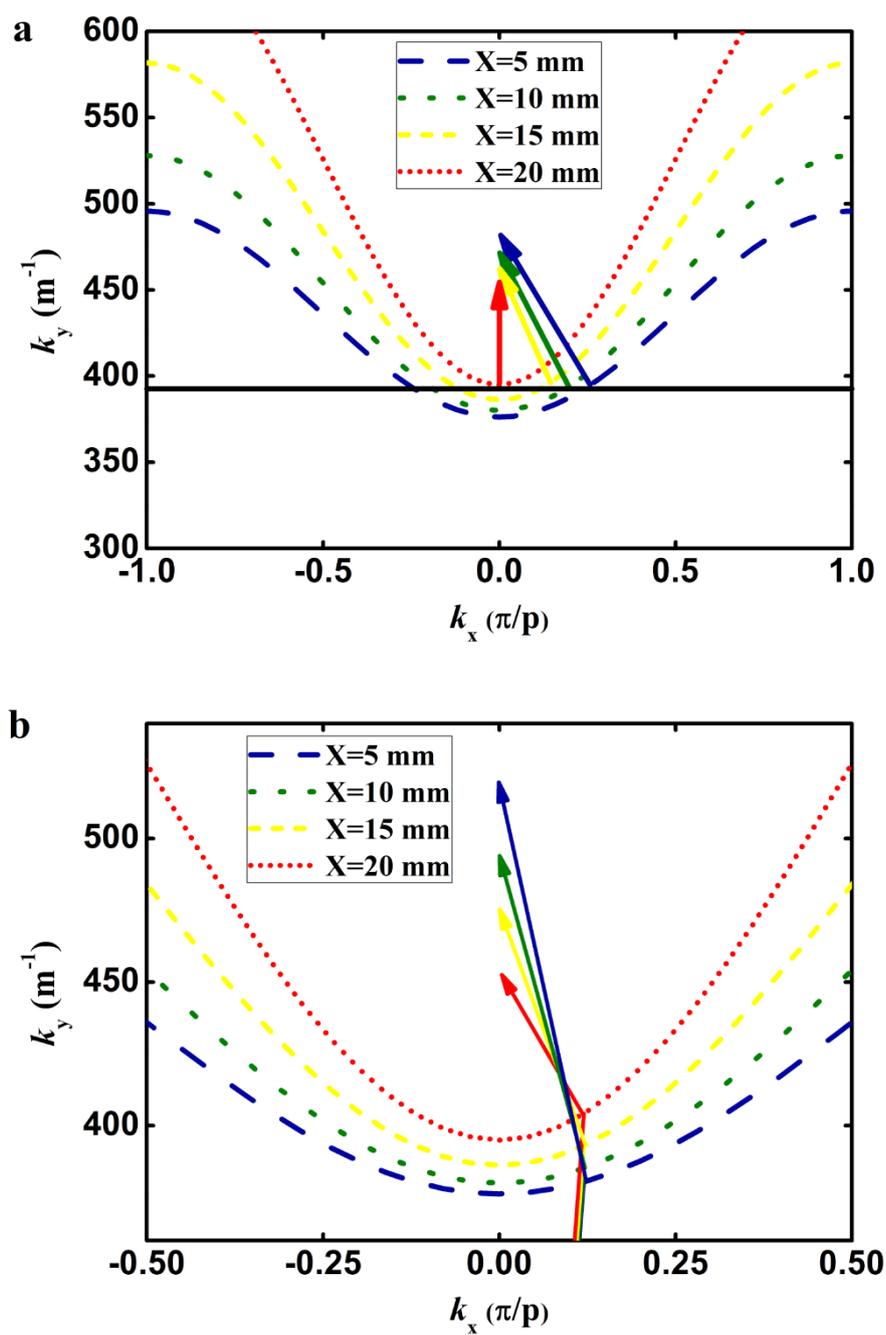

**Figure 5 | Dispersion relations of spoof SPPs.** (a) Local dispersion relation of spoof SPPs at the

different positions of MWGA for certain $k_y$ and (b) with the same lateral components.

**Fig. 6**

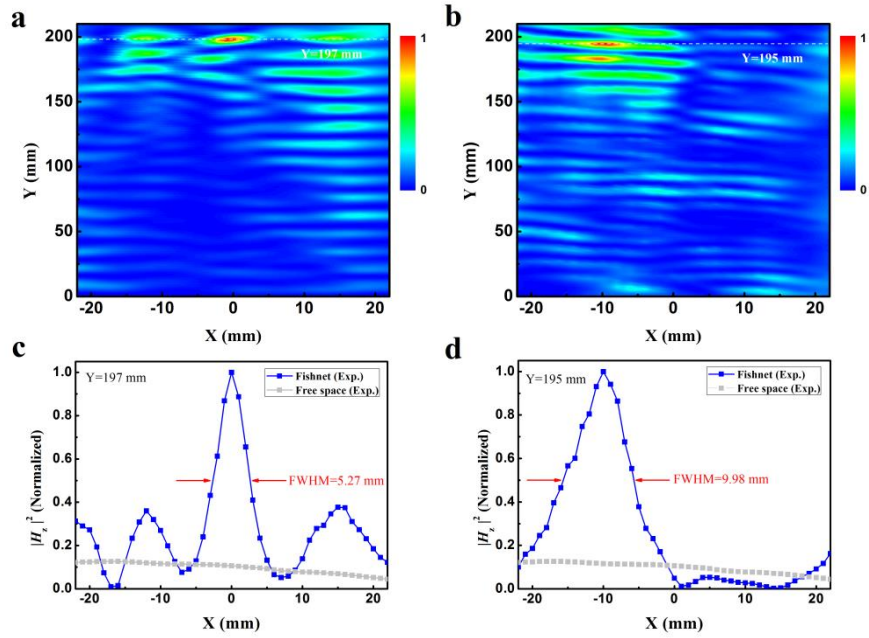

**Figure 6 | Experimental observation of subwavelength focusing.** Measured spatial distribution (a) /theta=0º, (b) theta=30º, and FWHM (c)/theta=0º, (d) theta=30º, of magnetic field intensity $|H_z|^2$ of stacked graded fishnet sample at frequency of 10.6GHz.